\documentclass{elsart}
\usepackage{amsfonts}
\usepackage{amssymb}
\usepackage{amsmath}
\usepackage[all]{xy}
\begin{document}

\begin{frontmatter}

\title{A note on unparticle in lower dimensions}

\author{Patricio Gaete\thanksref{cile}}
\thanks[cile]{e-mail address: patricio.gaete@usm.cl}
\address{Departamento de F\'{\i}sica, Universidad T\'ecnica
Federico Santa Mar\'{\i}a, Valpara\'{\i}so, Chile}

\author{Euro Spallucci\thanksref{infn}}
\thanks[infn]{e-mail address: spallucci@ts.infn.it }
\address{Dipartimento di Fisica Teorica, Universit\`a di Trieste
and INFN, Sezione di Trieste, Italy}

\begin{abstract}
Using the gauge-invariant but path-dependent variables formalism, we
examine the effect of the space-time dimensionality on a physical
observable in the unparticle scenario. We explicitly show that
long-range forces between particles mediated by unparticles are
still present whenever we go over into lower dimensions.
\end{abstract}
\end{frontmatter}

The physical consequences of nontrivial scale invariance in
high-energy interactions have been extensively discussed at the
present time
\cite{Strassler:2008bv,Liao:2007ic,Liao:2007fv,Liao:2007bx,Rizzo:2007xr},
\cite{Cheung:2007ap,Bander:2007nd,Cheung:2007jb,Cheung:2007zza},
\cite{Stephanov:2007ry,Kazakov:2007fn,Kazakov:2007su,Lee:2007xd,Lenz:2007nj}.
As is well known, the interest in studying this new scale invariant sector
is mainly due to the possibility of obtaining unusual properties of
matter with non-trivial scale invariance occurring in the infrared
regime \cite{Banks:1981nn}. This new sector has been called as the
unparticle sector \cite{Georgi:2007ek,Georgi:2007si}. We further
note that recently a novel way to describe unparticles has been
considered \cite{Krasnikov:2007fs}. The crucial ingredient of this
development is to introduce continuous mass spectrum objects, which
permits to interpret unparticle as a field with continuously
distributed mass.

In this context it may be recalled that one of the most interesting
of the phenomena predicted by unparticle physics is the existence of
long-range forces between particles mediated by unparticles
\cite{Liao:2007ic,Goldberg:2007tt}. More specifically, it was shown
that the corresponding modified Coulomb potential may be written as
\begin{equation}
V = \left( { - \frac{{q^2 }}{{4\pi }}} \right) \frac{1}{L} \left[ {1
+ \frac{2}{{\pi ^{2d_U  - 1} }}\frac{{\Gamma \left( {d_U  +
{\raise0.7ex\hbox{$1$} \!\mathord{\left/
 {\vphantom {1 2}}\right.\kern-\nulldelimiterspace}
\!\lower0.7ex\hbox{$2$}}} \right)\Gamma \left( {d_U  -
{\raise0.7ex\hbox{$1$} \!\mathord{\left/
 {\vphantom {1 2}}\right.\kern-\nulldelimiterspace}
\!\lower0.7ex\hbox{$2$}}} \right)}}{{\Gamma \left( {2d_U }
\right)}}\left( {\frac{l}{L}} \right)^{2d_U  - 2} } \right],
\label{Du01}
\end{equation}
where $l$ is a scale factor, given by $ l = \left(
{2^{\frac{1}{{2d_U  - 2}}} \Lambda _U } \right)^{ - 1} $. Here $d_U$
is a non-integral scale dimension of the unparticle field, and
$\Lambda_U$ defines a critical energy scale where the standard model
particles can interact with unparticles. We further note that a
different method for arriving at the same static potential profile
$(\ref{Du01})$, based on the gauge-invariant but path-dependent
variables formalism, was also developed in \cite{Gaete:2008wg}. An
important feature of this methodology is that it provides a
physically-based alternative to the usual Wilson loop approach. It
is interesting to observe that from expression $(\ref{Du01})$, when
$d_U=0$, the Coulomb potential correction became linear leading to
the confinement of static charges. It should, however, be noted that
the $d_U=0$ case is not allowed because the gamma function is not
analytic for this case. It is worth mentioning at this stage that a
range for $d_U$ of $1 < d_U \leq2$ has been considered in the
literature. In this connection it becomes of interest, in
particular, to recall that for $d_U < 1$ there is a nonintegrable
singularity in the differential decay rate into unparticles as
$E_U\longrightarrow0$ \cite{Georgi:2007ek}. As was observed by
Georgi \cite{Georgi:2007ek}, this is in accord with a theorem due to
Mack \cite{Mack} where it is shown that in an unitary theory fields
with $d_U < 1$ are not allowed. Thus, from a physical point of view,
the above remark ($d_U =0$ case) on the static potential
$(\ref{Du01})$ may be considered as another manifestation of the
arguments claimed in \cite{Georgi:2007ek}. Here it is important to
emphasize that the foregoing observations are restricted to three
space dimensions only, and it naturally raises the question of its
generalization in lower dimensions. In fact, it is not quite evident
that the same phenomenon will be repeated in two and one space
dimensions. The present work specifically deals with this problem,
where we examine the effect of the space-time dimensionality of the
problem under consideration on a physical observable.

It is worth recalling at this point that two-dimensional models have
been an extraordinary theoretical laboratory to test ideas in
quantum field theory.  Of particular interest are non-perturbative
issues like confinement and spectrum of models. Of these, the
Schwinger model \cite{Schwinger:1962tp} has probably enjoyed the
greatest popularity due to several features that it possesses. For
example, the spectrum contains a massive mode, the charge is
screened and confinement is satisfactorily addressed
\cite{Gross,GaeteSchmidt}. We further note that recently the
unparticle stuff in one space dimension has been studied
\cite{Georgi:2008pq}. In particular it was considered the
Sommerfield model \cite{Sommerfield}, that is, the exactly soluble
two-dimensional theory of a massless fermion coupled to a massive
vector boson. Notice that this model is the Schwinger model with an
additional mass term for the vector boson. As was explained in
\cite{Georgi:2008pq}, the Sommerfield model is an interesting analog
of a Banks- Zaks model \cite{Banks:1981nn}, approaching a free
theory at high energies and a scale invariant theory with
non-trivial anomalous dimensions at low energies. It is worth
recalling at this stage that Banks and Zaks investigated the unusual
properties of matter with non-trivial scale invariance in infra-red
regime. Interestingly, this new kind of stuff has no definite mass
at all. As mentioned before, this new sector has been called as the
unparticle sector \cite{Georgi:2007ek,Georgi:2007si}. The above
remark opens up the way to a stimulating discussion on the existence
of long-range forces between particles mediated by unparticles in the two-dimensional case. On the other hand, also it is important to recall here that
three-dimensional theories have been studied by various authors in
the last few years \cite{Deser:1981wh,Dunne:1998qy,Khare}.
As is well known, they are interesting because of its connection to the
high-temperature limit of four-dimensional theories
 \cite{Appelquist,Jackiw,Das} as well as for their applications to condensed
matter physics \cite{Stone}. Thus, as already mentioned, the main purpose
here is to examine the effects of the  space-time dimensionality on a
physical observable for the three and two-dimensional cases. To do
this, we will work out the static potential for the theories under
consideration by using the gauge-invariant but path-dependent
variables formalism along the lines of Ref. \cite{Gaete:2008wg}. As
a result, there are two generic features that are common in the
four-dimensional case and its lower dimensional extensions studied
here. First one,  the existence of long-range forces between
particles mediated by unparticles. The second point is related to
that in an unitary theory fields with $d_U < 1$ are not allowed.

We turn now to the problem of obtaining the interaction energy
between static point-like sources for the lower dimensional
extensions under consideration. To do this, we shall compute the
expectation value of the energy operator $H$ in the physical state
$|\Phi\rangle$ describing the sources, which we will denote by $
{\langle H\rangle}_\Phi$. In order to introduce some notation for
our subsequent work we start from the four-dimensional space-time
Lagrangian density \cite{Krasnikov:2007fs}:
\begin{equation}
\mathcal{L} = \sum\limits_{k = 1}^N {\left[ { - \frac{1}{{4e_k^2 }}
F^{k\mu \nu } F_{\mu \nu }^k  + \frac{{m_k^2 }}{{2e_k^2 }}\left(
{A_\mu ^k  - \partial _\mu  \varphi ^{k} } \right)^2 } \right]},
\label{Du05}
\end{equation}
where $m_{k}$ is the mass for the $k-th$ scalar field. Following our
earlier procedure \cite{Gaete:2008wg}, to compute the interaction
energy we need to carry out the integration over the
$\varphi$-fields. Once this is done, we arrive at the following
effective theory for the gauge fields:
\begin{equation}
\mathcal{ L}_{eff}^{(3+1)}   = \sum\limits_{k = 1}^N {\frac{1}{{e_k^2
}}} \left[ { - \frac{1}{4}F_{\mu \nu }^k \left( {1 + \frac{{m_k^2
}}{\Delta_{(3+1)} }} \right)F^{k\mu \nu } } \right]. \label{Du10}
\end{equation}

Next, in order to obtain the corresponding effective Lagrangian
density in $(2+1)$ dimensions, we compactify one spacelike dimension
by using a sort of Kaluza-Klein approach \cite{Gaete:2007ry}. It
follows that the expression $(\ref{Du10})$ can be rewritten as
\begin{equation}
\mathcal{ L}_{eff}^{KK}   = \sum\limits_{k = 1}^N {\sum\limits_{n =
0}^\infty  {\frac{1}{{e_k^2 }}} } \left[ { - \frac{1}{4}F_{\mu \nu
}^k \left( {1 + \frac{{m_k^2 }}{{\Delta _{\left( {2 + 1} \right)}  +
a^2_n }}} \right)F^{k\mu \nu } } \right] , \label{Du15}
\end{equation}
with $a^2_n  \equiv {\raise0.7ex\hbox{${n^2 }$} \!\mathord{\left/
 {\vphantom {{n^2 } {R^2 }}}\right.\kern-\nulldelimiterspace}
\!\lower0.7ex\hbox{${R^2 }$}}$ , and $R$ is the compactification
radius.
In the limit $R\to\infty$ the difference between nearby energy
levels vanish and the $3+1$ dimensional continuum spectrum is
recovered. In the opposite case, $R\to 0$, the massive modes becomes
more and more heavy and decouple from the physical spectrum. The
only surviving mode corresponds to $n=0$ and describes the
dimensionally reduced model in $2+1$ dimensions.
With this at hand, we can now compute the interaction energy for a
single mode in Eq. $(\ref{Du15})$. The canonical Hamiltonian can be
worked as usual and is given by
\begin{equation}
H_C  = \int {d^2 } x\left\{ { - \frac{1}{2}\Pi ^i \left( {1 +
\frac{{m_{k}^2 }}{{\Delta } + a^2_n}} \right)^{ - 1} \Pi _i  + \Pi ^i
\partial _i A_0  + \frac{1}{4}F_{ij} \left( {1 + \frac{{m_{k}^2
}}{{\Delta }+ a^2_n}} \right)F^{ij} } \right\}, \label{Du20}
\end{equation}

where $\Pi ^{i}  =  - \left( {1 + \frac{{m_{k}^2 }}{\Delta + a_{n}^2
}} \right)F^{k0i }$ are the canonical momenta. Here, we have
simplified our notation by setting $\Delta_{(2+1)}\equiv\Delta$.

Following our earlier discussion \cite{Gaete:2008wg}, the resulting
static potential for two opposite charges located at ${\bf y}$ and
${\bf y^{\prime}}$ takes the form:
\begin{equation}
V = \sum\limits_{k = 1}^N  {\sum\limits_{n = 0}^\infty
{\frac{1}{{e_k^2 }}} } \left\{ { - \frac{{q^2 }}{{2\pi }}K_0 \left(
{M_{k,n} L} \right) + \frac{{q^2 a^2_n }}{{4M_{k,n}}}L} \right\}, \label{Du25}
\end{equation}
where $M^2_{k,n}  \equiv m_k^2  + a^2_n$, $L\equiv|{\bf y}-{\bf
{y^\prime}}|$ and $K_0 \left( {M_{k,n}L} \right)$ is a modified Bessel
function.

In Eq.(\ref{Du25}) the first terms is the one which will account for
the ``un-particle'' corrections, while the second one is the
``Coulombic interaction'' in $2+1$ dimensions. It is worth to remark
contrary to expectation it is not a logarithmic potential, rather we
find that massive KK-modes produce a linear term. This difference
can be traced back to the different definition of static potential.
Instead of using the standard one, we use the gauge
invariant/path-dependent approach developed in \cite{GaeteSchmidt,Gaete:1998vr}.
However, the sum over $n$ is ill-defined as it is linearly divergent.
The ``string-tension''  $\sigma_k$ for each field
is given by the divergent sum
\begin{equation}
\sigma_k  = \frac{{q^2 }}{4} \frac{1}{R}\sum\limits_{n=0}^{\infty}
{\frac{{n^2 }}{{\sqrt {n^2  + m_k^2 R^2 } }}}. \label{Du30}
\end{equation}
and needs some regularization prescription. It follows that this
part of the result is necessarily ambiguous and cannot be taken too
seriously as a candidate to a confinement potential. As an example,
one can see that in the decoupling limit $m_kR<< 1$, zeta-function
regularization
gives a negative string tension.\\
Since our main motivation is to compute the correction to the static
potential for the three-dimensional case, we drop out this
term and consider the zero mode only. Thus, it follows that
\begin{equation}
V_{n=0} =  - \frac{{q^2 }}{{2\pi }}\sum\limits_{k = 1}^N {\frac{1}{{e_k^2
}}} K_0 \left( {\sqrt {m_k^2 } L} \right). \label{Du35}
\end{equation}
Before switching-on un-particle effects, we introduce a further
simplification by assuming that the test charges are at distance $L>
1/m_k$ for any $k$. As we are looking for long-range forces this is
a fairly reasonable choice. In this case, the Bessel function can be
approximated with  $exp(- {\sqrt {m_k^2 } L})$ and the static
potential  (\ref{Du35}) may be written as
\begin{equation}
V_{n=0} \simeq  - \frac{{q^2 }}{{2\pi }}\sum\limits_{k = 1}^N {\frac{1}{{e_k^2
}}} e^{ - m_k L} . \label{Du40}
\end{equation}
 Now we are ready to include unparticle effects in the potential
 (\ref{Du40}).
By following \cite{Krasnikov:2007fs} and
\cite{Gaete:2008wg}, we go into the continuum mass spectrum
limit, $N \to\infty$,  and replace the sum over $k$
by an integral
\begin{equation}
V_{n=0} \to V_U=\left( { - \frac{{q^2 }}{{2\pi e^2 }}}
\right)\frac{A_{d_U}}{\Lambda_U^{2d_U -2}}
\int_0^\infty  {t^{d_U  - 2} e^{ - \sqrt t L} } dt,
\label{Du45}
\end{equation}
where $t=m_{k}^{2}$,  $\rho\left(\,t\,\right)\equiv
{t^{d_U  - 2} }$ is the spectral density,
and $A_{d_{U}}$ is a normalization factor which is given by
\begin{equation}
A_{d_U }  \equiv \frac{16\pi^{5/2}}{\left(\, 2\pi\,\right)^{2d_U}}
\frac{{\Gamma \left( {d_U  + {\raise0.7ex\hbox{$1$} \!\mathord{\left/
 {\vphantom {1 2}}\right.\kern-\nulldelimiterspace}
\!\lower0.7ex\hbox{$2$}}} \right)}}{{\Gamma \left( {d_U  - 1} \right)
\Gamma \left( {2d_U } \right)}}, \label{Du50}
\end{equation}
where $d_{U}$ is the scale dimension of the unparticle field. We also
note here that in Eq. (\ref{Du45}) we have assumed
 $e^2=e_k^2$. A direct computation on the $t$-variable yield
\begin{equation}
V_U = \left( { - \frac{{q^2 }}{{2\pi e^2 l}}} \right)A_{d_U } \sqrt {2\pi }
\Gamma \left( {2d_U  - {\raise0.7ex\hbox{$5$} \!\mathord{\left/
 {\vphantom {5 2}}\right.\kern-\nulldelimiterspace}
\!\lower0.7ex\hbox{$2$}}} \right)\left( {\frac{l}{L}} \right)^{2d_U
- 2}. \label{Du55}
\end{equation}
Using (\ref{Du50}), we see that $V_U$ reads
\begin{equation}
V_U = \left( { - \frac{{q^2 }}{{2\pi e^2 l}}} \right)\frac{1}{{\left( \pi
\right)^{2d_U  - 3} }}\frac{1}{{\left( 2 \right)^{2d_U  - {9
\mathord{\left/ {\vphantom {9 2}} \right. \kern-\nulldelimiterspace}
2}} }}\frac{{\Gamma \left( {d_U  + {1 \mathord{\left/ {\vphantom {1
2}} \right. \kern-\nulldelimiterspace} 2}} \right)\Gamma \left(
{2d_U  - {5 \mathord{\left/ {\vphantom {5 2}} \right.
\kern-\nulldelimiterspace} 2}} \right)}}{{\Gamma \left( {d_U  - 1}
\right)\Gamma \left( {2d_U } \right)}}\left( {\frac{l}{L}}
\right)^{2d_U  - 2}, \label{Du55bis}
\end{equation}
where we have introduced the scale factor $l \equiv \frac{1}
{\Lambda _U }$. It may be noted that for  $d_U= \frac{1}{2}$,
expression (\ref{Du55bis}) reduces to
\begin{equation}
V_U = \left( {\frac{{q^2 }}{{2\pi e^2 l}}} \right)\frac{{\pi ^2 2^{{9
\mathord{\left/ {\vphantom {9 2}} \right. \kern-\nulldelimiterspace}
2}} }}{{3l}}L. \label{Du56}
\end{equation}
From this, one infers the key role played by the scale dimension
($d_U$) in transforming the long-range potential into the confining
one. In this way, for the three-dimensional case, unparticles with
$d_U < 1$ would be allowed.  However, by unitarity considerations
the obstruction for unparticles ($d_U < 1$) present in  the
four-dimensional case is still present whenever we go over into
three dimensions.

The $(1+1)$-dimensional case may be studied in the same way as we
did in the $(2+1)$- dimensional counterpart. In such a case, the
theory under consideration is given by
\begin{equation}
\mathcal{ L} = {\sum\limits_{k = 1}^N \sum\limits_{n,m = 0}^\infty
{\frac{1}{{e_k^2 }}} } \left\{ { - \frac{1}{4}F_{\mu \nu }^k \left(
{1 + \frac{{m_k^2 }}{{\Delta  + \kappa_{n,m} ^2 }}} \right)F^{k\mu
\nu } } \right\}, \label{Du60}
\end{equation}
where $\kappa_{n,m} ^2  = {\raise0.7ex\hbox{${n^2 }$} \!\mathord{\left/
 {\vphantom {{n^2 } {R_1^2 }}}\right.\kern-\nulldelimiterspace}
\!\lower0.7ex\hbox{${R_1^2 }$}} + {\raise0.7ex\hbox{${m^2 }$}
\!\mathord{\left/ {\vphantom {{m^2 } {R_2^2
}}}\right.\kern-\nulldelimiterspace} \!\lower0.7ex\hbox{${R_2^2
}$}}$. The situation here is analogous to that encountered in the
Schwinger model \cite{GaeteSchmidt}.  This allows us to write the
static potential as
\begin{equation}
V =  {\sum\limits_{k = 1}^N \sum\limits_{n,m = 0}^\infty
{\frac{1}{{e_k^2 }}} } \left\{ {\frac{{q^2 }}{{2\lambda }}\left( {1
+ \frac{{\kappa_{n,m} ^2 }}{{\lambda ^2 }}} \right)\left( {1 - e^{ -
\lambda L} } \right) + \frac{{q^2 }}{2}\frac{{\kappa_{n,m} ^2
}}{{\lambda ^2 }}L} \right\}, \label{Du65}
\end{equation}
where $\lambda ^2  \equiv m_k^2  + \kappa _{n,m}^2$. Since we are
interested in estimating the long-range correction to the static
potential, we will retain only the zero mode contribution in the
expression (\ref{Du65}). Thus the static potential simplifies to
\begin{equation}
V_{0,0} = \left( { - \frac{{q^2 }}{2}} \right)\sum\limits_{k = 1}^N
{\frac{1}{{e_k^2 }}} \frac{{e^{ - \sqrt {m_k^2 } L} }}{{\sqrt {m_k^2
} }}. \label{Du70}
\end{equation}

Following our earlier procedure, we see that the unparticle potential
corresponding to (\ref{Du70}) takes the form

\begin{equation}
V_U = \left( { - \frac{{q^2 }}{2 e^2 l}} \right)\frac{{\pi ^{\frac{5}{2} -
2d_U } }}{{2^{2d_U  - 5} }}\frac{{\Gamma \left( {d_U  +
{\raise0.7ex\hbox{$1$} \!\mathord{\left/ {\vphantom {1
2}}\right.\kern-\nulldelimiterspace} \!\lower0.7ex\hbox{$2$}}}
\right)\Gamma \left( {2d_U  - 3} \right)}}{{\Gamma \left( {d_U  - 1}
\right)\Gamma \left( {2d_U } \right)}} 
\left( {\frac{l}{L}} \right)^{2d_U  - 3} . \label{Du75}
\end{equation}

Hence we see that for $d_U=1$, the potential has a linear dependence
from the distance $L$. However, the string tension has to evaluated carefully
because of the simple poles in the gamma functions for $d_U=1$. Thus, we
define $V_U$ in the limit $d_U\to 1$ as

\begin{equation}
\lim_{d_U\to 1} V_U =\lim_{\epsilon\to 0}
\left( { - \frac{{q^2 }}{2 e^2 l}} \right)\frac{{\pi ^{\frac{5}{2} -
2d_U } }}{{2^{2d_U  - 5} }}\frac{{\Gamma \left( {d_U  +
{\raise0.7ex\hbox{$1$} \!\mathord{\left/ {\vphantom {1
2}}\right.\kern-\nulldelimiterspace} \!\lower0.7ex\hbox{$2$}}}
\right)\Gamma \left( {2+ 2\epsilon  - 3} \right)}}{{\Gamma \left(
{1+\epsilon  - 1}\right)\Gamma \left( {2d_U } \right)}}
\left( {\frac{l}{L}} \right)^{2d_U  - 3}
\end{equation}

In the limit $\epsilon \to 0$ the poles in the two Gamma function
cancel leading to a finite result:

\begin{equation}
\lim_{\epsilon\to 0}\frac{\Gamma \left(\, -1+ 2\epsilon\,\right)}
{\Gamma \left(\, \epsilon\,\right)}= -\frac{1}{2}\ .
\end{equation}

Thus, in $1+1$ dimension we recover the correct confining potential

\begin{equation}
V_U^{d_U=1}=\sigma\, L \ ,\qquad \sigma = \
\frac{{2\pi q^2 }}{{e^2 l^2 }}\ .
\end{equation}

From the above result (\ref{Du75}) it is meaningful to ask whether a
similar thing happens in the case of the Sommerfield model studied
in \cite{Georgi:2008pq}. The Sommerfield model  is a Lagrangian
field theory which describes massless spinors interacting with a massive
vector boson, in $1+1$ dimensions. The Lagrangian density reads:
\begin{equation}
\mathcal{ L} = \bar \psi  \left( {i{\slash\! \! \! \partial}  - e
{\slash\! \! \! \! A} } \right)\psi - \frac{1}{4}F_{\mu \nu } F^{\mu
\nu }  + \frac{{m_0^2 }}{2}A_\mu A^\mu, \label{Du80}
\end{equation}
where $m_{0}$ is the mass for the vector boson $A_{\mu}$. 
In \cite{Georgi:2008pq} authors study the transition between
unparticle behavior at low energy and free particle behavior 
at high energy from the vantage point of the exactly solvable
model (\ref{Du80}) mimicking Banks-Zacks model in lower dimensions. 
Here, we are interested to determine the interaction
energy in the unparticle phase and  the eventual presence
of a linear confining term. \\ 
Following our earlier adaptation of the Stueckelberg procedure \cite{Gaete:2008wg}, we first restore
gauge invariance by means of a suitable compensating field for the sake of consistency
with our gauge invariant definition of interaction potential. Then, 
 we integrate out both fermions and compensator field to obtain an effective theory for the 
 gauge vector $A_{\mu}$.
Once this is done, we arrive at the following effective Lagrangian
density:
\begin{equation}
\mathcal{ L} =  - \frac{1}{4}F_{\mu \nu } \left( {1 + \frac{{m^2
}}{\Delta }} \right)F^{\mu \nu }, \label{Du85}
\end{equation}
where $m^2  = m_0^2  + {\raise0.7ex\hbox{${e^2 }$} \!\mathord{\left/
 {\vphantom {{e^2 } \pi }}\right.\kern-\nulldelimiterspace}
\!\lower0.7ex\hbox{$\pi $}}$. As a consequence, the static potential
is given by \cite{GaeteSchmidt}:
\begin{equation}
V = \left( { - \frac{{q^2 }}{2}} \right)\frac{{e^{ - mL} }}{m}.
\label{Du85bis}
\end{equation}
Again, by considering unparticle as a field with continuously
distributed mass, we can write Eq. (\ref{Du85bis}) as
\begin{equation}
V = \left( { - \frac{{q^2 }}{2}} \right)\sum\limits_{k = 1}^N
{\frac{{e^{ - \sqrt {m_k^2 } L} }}{{\sqrt {m_k^2 } }}}. \label{Du90}
\end{equation}
In the same way as was done in the previous case, one finds
\begin{equation}
V_U = \left( { - \frac{{q^2 }}{2}} \right)A_{d_U } \Gamma \left( {2d_U
- 3} \right)2l  \left( {\frac{l}{L}} \right)^{2d_U  - 3} . \label{Du95}
\end{equation}
in agreement with Eq. (\ref{Du75}). In this way, both the Proca-Maxwell
and the Sommerfield model with continuously
distributed mass leads to the same static potential.\\

Acknowledgments.\\
We would like to thank the referee of our previous paper
\cite{Gaete:2008wg} for inspiring the present work. P. G. was
partially supported by Fondecyt (Chile) grant 1080260.

\end{document}